\documentclass{iopart}
\usepackage{iopams}
\usepackage{graphicx}
\usepackage{color}

\begin{document}

\title[Design of a speed meter interferometer proof-of-principle 
experiment]{Design of a speed meter interferometer proof-of-principle 
experiment}

\author{\mbox{C~Gr\"{a}f}\hspace{.3mm}$^{1}$, \mbox{B W Barr}$^{1}$, 
\mbox{A S Bell}$^{1}$, \mbox{F Campbell}$^{1}$, \mbox{A V Cumming}$^{1}$, 
\mbox{S L Danilishin}$^{2}$, 
\mbox{N A Gordon}$^{1}$, \mbox{G D Hammond}$^{1}$, \mbox{J Hennig}$^{1}$, 
\mbox{E A Houston}$^{1}$, \mbox{S H Huttner}$^{1}$, \mbox{R A Jones}$^{1}$, 
\mbox{S S Leavey}$^{1}$, \mbox{H L\"uck}$^{3}$, \mbox{J Macarthur}$^{1}$, 
\mbox{M Marwick}$^{1}$, \mbox{S Rigby}$^{1}$, 
\mbox{R Schilling}$^{3}$, \mbox{B 
Sorazu}$^{1}$, \mbox{A Spencer}$^{1}$,
\mbox{S Steinlechner}$^{1}$, \mbox{K A Strain}$^{1}$ and \mbox{S Hild}$^{1}$}
\ead{christian.graef@glasgow.ac.uk}
\vskip 1mm
\address{$^{1}$\,SUPA, School of Physics and Astronomy, The University 
of Glasgow, Glasgow, G12\,8QQ, UK}
\address{$^{2}$\,School of Physics, University of Western Australia, 35 
Stirling Hwy, Crawley 6009, Australia}
\address{$^{3}$\,Max--Planck--Institut f\"{u}r Gravitationsphysik 
(Albert-Einstein-Institut) and Leibniz Universit\"{a}t Hannover, Callinstr.~38, 
D-30167 Hannover, Germany}

\begin{abstract}
The second generation of large scale interferometric gravitational wave 
detectors will
be limited by quantum noise over a wide frequency range in their detection 
band.
Further sensitivity 
improvements for future upgrades or new detectors beyond the second generation 
motivate the development of measurement schemes to mitigate the impact of 
quantum noise in these instruments. 
Two strands of development are being pursued to reach this goal, focusing 
both on modifications of the well-established Michelson detector 
configuration and development of different detector topologies.
In this paper, we present the design of the world's first Sagnac speed meter 
interferometer which is currently being constructed at the University of 
Glasgow. With this proof-of-principle experiment we aim to demonstrate the 
theoretically predicted lower quantum noise in a Sagnac interferometer 
compared to an equivalent Michelson interferometer, to qualify Sagnac 
speed meters for further research towards an implementation in a future 
generation large scale gravitational wave detector, such as the planned 
Einstein Telescope observatory.

\end{abstract}

\pacs{04.80.Nn, 07.60.Ly, 42.50.Lc}


\section{Introduction}\label{sec:intro}

The world-wide network of interferometric gravitational wave (GW) detectors is 
currently in an 
upgrade phase towards what is commonly referred to as the \emph{second 
generation} of detectors. This detector 
network is formed of the Advanced LIGO detectors (USA) 
\cite{aLIGO}, the Advanced Virgo detector (Italy/France) \cite{adVirgo}, the 
GEO-HF detector (Germany/UK) \cite{GEO-HF} and the new Japanese KAGRA 
observatory \cite{KAGRA}. All of these detectors are based on 
the classical Michelson topology and 
will be limited in their sensitivities by quantum noise over a large range of 
frequencies. Quantum noise originates from the quantum 
fluctuations of the laser light, which is used for the measurement of GW 
induced strain, and manifests as quantum radiation pressure noise (QRPN) and 
shot noise, dominating at low and high frequencies, respectively.

\begin{figure}[t]
\centering
\includegraphics[width=0.8\textwidth]{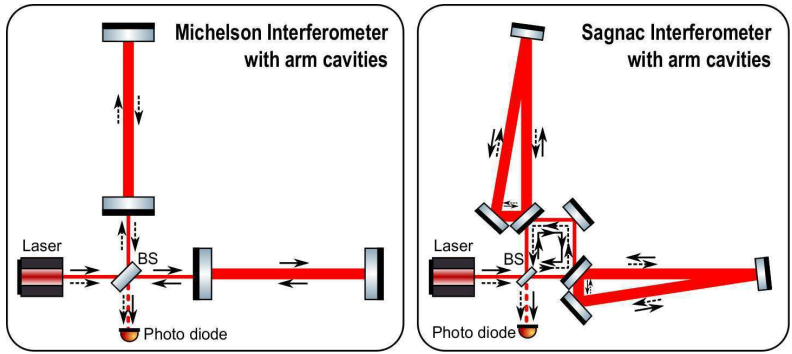}
\caption{Comparison of Michelson with Sagnac interferometer topologies, 
with arm cavities in both cases. In the Michelson topology the positions 
of the mirrors in the two arms of the 
interferometer are sensed either by the beam transmitted or the beam reflected 
at the beam splitter. However, in the Sagnac interferometer both 
beams, travelling in the clockwise and the counter clockwise direction, sense 
the relative positions of all mirrors, though at different times. Consequently, 
the resulting signal extracted from the main photo detector in the Sagnac 
interferometer carries information about the relative velocity of the test mass 
mirrors.}
\label{fig:MI_vs_SAG_layout}
\end{figure}

Michelson-based laser interferometers, {\em cf.~}figure 
\ref{fig:MI_vs_SAG_layout}, 
are so called \emph{position meters}, i.e.~the measured observable is the 
position of the test mass mirrors which does not commute with itself at 
subsequent moments of time. This non vanishing commutator yields an uncertainty 
relation for subsequent position measurements which 
enforces the so-called \emph{Standard Quantum Limit} (SQL). 
The notion of the SQL was originally coined by Braginsky \cite{Braginsky92} and 
was believed to set a limit to the achievable sensitivity of a precision 
measurement apparatus.
It was later 
realised by Braginsky 
that it is in fact possible to overcome the SQL by performing a so-called 
\emph{Quantum Non-Demolition} (QND) measurement 
\cite{Braginsky80}, e.g.~by introducing cross-correlations in the quantum 
noise in an interferometer.

To achieve sensitivities an order 
of magnitude higher than in the second generation GW observatories 
\cite{Punturo2010b} requires quantum noise to be reduced. In particular, the 
characteristic $ \propto 1/f^{2}$ frequency dependence of QRPN causes it to 
occlude the low-frequency band which contains a rich population of potential 
sources \cite{Sathya2012}.
An important cornerstone in the designs of GW detectors beyond the second 
generation, such as the planned \emph{Einstein Telescope} detector 
\cite{Punturo2010a}, is the use 
of \emph{frequency dependent squeezed vacuum states of light} \cite{Hild2011}. 
This technique was, among others, proposed by Kimble {\em et al.~}\cite{KLMTV} 
and aims for the conversion of conventional Michelson-based GW detectors into 
QND interferometers, to achieve a broadband improvement of the quantum noise 
limited sensitivity. However, all the 
schemes proposed in \cite{KLMTV} have in common that they come at the expense 
of substantially increased technical complexity as they require the adoption of 
large scale low-loss filter cavities.

A new approach which is based on performing a measurement of an 
observable which is intrinsically a QND 
observable, i.e.~a conserved quantity, was first introduced to the field of GW 
interferometry by Braginsky.
In so-called \emph{speed 
meters} the intention was to measure the momentum of the test masses, for which 
their relative velocity was thought to be a good proxy
\cite{Braginsky90}.
However, a more thorough analysis has shown that in an opto-mechanical system, 
in which test mass dynamics and light fields are coupled, velocity is not a 
conserved quantity (see section 4.5.2 of \cite{DanKhaLRR2012} and references 
therein) and therefore is also limited by the SQL. Nevertheless, the QRPN of 
the speed meter is naturally reduced. This opens the way for a broadband 
beating of the SQL using homodyne readout of the optimally chosen light 
quadrature at the dark port.

The first practical speed meter configurations were based on the concept of 
\emph{signal sloshing} \cite{Braginsky00} and aimed at modifications of the 
optical layouts of the already well-established Michelson-based 
configurations, by introducing so-called \emph{sloshing cavities} in the 
Michelson output 
port \cite{Purdue02}.
In 2003, Chen pointed out that a Sagnac interferometer, 
{\em cf.~}figure \ref{fig:MI_vs_SAG_layout}, is a speed meter \emph{per se}, as 
the 
signal exiting the interferometer is proportional to the 
time-dependent variation of the relative test mass positions, i.e.~to their 
relative velocity \cite{Chen2003}. This makes Sagnac interferometers 
the more favourable speed meter configuration since no additional large 
scale cavity needs to be included in the optical layout for which 
ultra-low loss operation and low noise feedback control need to be pioneered.

Sagnac interferometers, in a zero area configuration, had already been studied 
as 
candidate configurations for GW detectors more than a decade ago, prior to the 
implementation of the first generation of large scale detectors 
\cite{Sun96, 
Shaddock98, Beyersdorf02}. However, these investigations did not reveal 
significant advantages over Michelson-based configurations because Michelson 
interferometers were far from being limited by the SQL and the quantum back 
action evasion potential of Sagnac-based configurations was unknown.

In the scope of the conceptual design study for the 
Einstein Telescope (ET) third generation GW observatory it was shown that, in 
theory, Sagnac speed meter interferometers outperform Michelson-based 
interferometers with comparable parameters in terms of higher broadband 
sensitivities \cite{MuellerEbhardt09}. This result was recently confirmed by 
Miao {\em et al.~}in \cite{Miao2014}.
Despite the advantages of speed meter configurations, Michelson-based 
interferometers were chosen as the baseline design for the ET 
detectors \cite{ETDesignStudy}. 
This choice was to a large extent motivated by the fact that Michelson-based 
interferometers have been highly refined and optimised 
for gravitational wave detection over a period of almost four decades.

In the longer term, Sagnac speed meter configurations are 
prospective candidates for replacing Michelson-based GW 
detectors. However, the superiority of Sagnac-based configurations is yet to be 
demonstrated experimentally. Hence, the main 
objectives of our Sagnac speed meter experiment are i) the realisation of an 
ultra-low noise, QRPN dominated speed meter test bed and 
ii) the experimental demonstration 
of reduced quantum back action noise in comparison to an equivalent 
Michelson-based interferometer. By reaching these objectives we aim for 
qualifying the Sagnac speed meter concept for further research towards an 
implementation in a future GW detector, such as the Einstein Telescope.

\section{Conceptual approach}\label{sec:idea}
In this section we illustrate the conceptual approach to the Glasgow Sagnac 
Speed 
Meter (SSM) experiment design. With this experiment we aim for the 
demonstration of 
reduced quantum back-action noise in a Sagnac interferometer 
in comparison to an equivalent Michelson interferometer. However, it 
must be noted that beating the SQL, as is aimed for in 
conceptually similar experiments, as e.g.~\cite{Dahl2012}, is not one of the 
explicit goals of our SSM experiment.

The starting point for the design of our SSM experiment was a 
conventional Michelson interferometer with Fabry-Perot cavities in the arms, 
with quantum noise dominating over the total classical noise in the 
instrument at low frequencies. The low-frequency sensitivity of this Michelson 
interferometer design then served as a benchmark for the SSM conceptual design, 
in the sense that a successful equivalent speed meter design needs to yield 
better low-frequency sensitivity than could be achieved with the reference 
Michelson interferometer design.

The main drivers in the conceptual design process were i) an 
optimisation of the design to enhance QRPN and ii) 
the reduction of classical noise sources which potentially mask quantum noise. 
To fulfill the former requirement, the ratio of the circulating light power and 
the arm cavity mirror 
masses was chosen, to push the frequency at which shot noise and QRPN are equal 
to above approximately 5\,kHz. Here, the 
challenge is to identify a set of design parameters 
resulting in sufficiently low classical noise while maintaining the technical 
feasibility and operability of the experiment. In our 
case this was achieved by choosing a high arm cavity finesse resulting in 
circulating powers in the kW range and by choosing a mirror mass for the arm 
cavity input couplers of approximately one gram. Regarding the test mass and 
intra-cavity power regimes, our SSM experiment is comparable e.g.~to the 
experiments described in \cite{Corbitt2007a, Corbitt2007b}. Generally, to keep 
the technical complexity of the SSM experiment at a manageable level, the 
design was to the largest possible extent based on well-established 
experimental techniques and the adoption of new techniques, i.e.~so far 
untested or immature ones, was avoided where possible.

In contrast to even lighter test 
masses, such as membranes or cantilevers, our cavity input mirrors have 
the advantage of lower thermal noise. Also, in 
this mass regime it is feasible to employ the same concepts of low-dissipation 
mirror suspension systems as well as actuation schemes as those used 
in prototype interferometer experiments and full-scale GW detectors. To isolate 
the experiment 
from seismic and environmental noise and, at the same time, keep 
thermal noise at a sufficiently low level, the in-vacuum optics will be 
suspended using cascaded pendulum suspensions. For the core interferometer 
optics these will feature multiple eddy-current damped pendulum stages with an 
all-monolithic final stage for horizontal isolation and cantilever-mounted 
blade 
springs for 
vertical isolation. Additional filtering of seismic disturbances will be 
provided by a passive stack-type pre-isolation system consisting of four 
alternating layers of stainless steel mass elements and fluoroelastomer spring 
elements. 

\begin{figure}[Ht]
\centering
\includegraphics[width=0.8445\textwidth]
{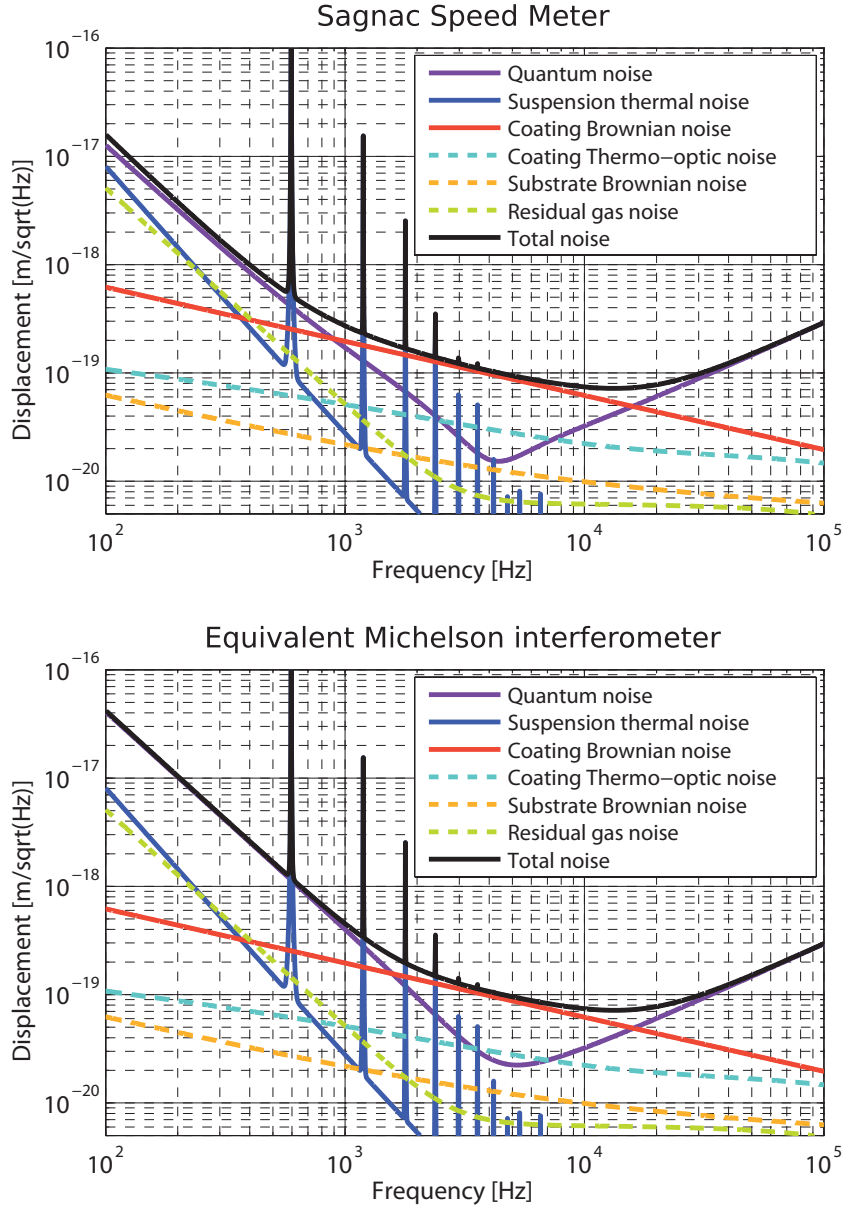}
\caption{\textbf{Top plot:} Top level displacement noise of the Glasgow 
SSM experiment. An input laser power of 300\,mW was assumed. The
optical round trip loss in the Fabry-Perot arm cavities was accounted for as 
25\,ppm. The main interferometer signal is read out by means of a balanced 
homodyne scheme with optimised homodyne angle.
\textbf{Bottom plot:} Displacement noise of an equivalent (in terms 
of effective mirror mass and circulating power in the arm cavities) 
Michelson-based 
interferometer. An identical input laser power of 300\,mW and an
optical round trip loss in the Fabry-Perot arm cavities of 25\,ppm were 
assumed. 
For the extraction of the main interferometer signal homodyne readout of the 
phase quadrature was assumed.}
\label{fig:SSM_MI_noise_budgets_combined_25ppm_loss}
\end{figure}

An arm cavity round trip length of approximately 2.83\,m was chosen. The short 
cavity length allows a compact 
physical 
arrangement designed to yield good common mode rejection of seismic 
noise.  It also provides low susceptibility to frequency noise, allowing 
a measurement sensitivity between $10^{-18}$\,m$/\sqrt{\textnormal{Hz}}$ and
$10^{-19}$\,m$/\sqrt{\textnormal{Hz}}$ at 
frequencies of a few hundreds of Hz. A selection of key design parameters of 
the 
SSM experiment is summarised in table \ref{table:ssm_design_parameters}.
\begin{table}[th]

\caption{\footnotesize\rm Key design 
parameters of 
the Glasgow SSM 
interferometer experiment. A more detailed parameter set will be developed in 
the course of the 
project. \label{table:ssm_design_parameters}}
\centering
\small
\begin{tabular}{p{4.5cm}|p{7.4cm}}
\hline
\textbf{Parameter} & \textbf{Value}\\
\hline\hline

Laser source & Nd:YAG NPRO laser, $2$\,W$ @ 1064$\,nm \\

Spatial mode filtering & Single mode optical fibres + triangular pre-mode 
cleaner cavity\\

\hline

Vacuum system & Two vacuum tanks, 1\,m diameter, 1.5\,m centre-to-centre 
length\\

Targeted pressure & $\approx10^{-6}$\,mbar, dominated by 
water vapour \\

Operating temperature & Room temperature\\

Lab environment & Class 1000 clean room by design -- better performance in 
measurements; additional clean room tents.\\

\hline

Seismic pre-isolation & Passive system with four stack layers, comprised 
of 
stainless steel mass elements and Fluorel$^\textnormal{\tiny{\textregistered}}$ 
springs \\

Core optics mass & Arm cavity incoupling mirrors $\approx 
1$\,g; arm cavity end mirrors $100$\,g\\

Core optics suspensions & Triple pendulum suspensions: steel wires in 
the upper 
stages, all monolithic final stage. Blade springs for vertical 
isolation. \\

Fused silica fibres for 1\,g mirrors& 20\,$\mu$m diameter, 5\,cm length\\

Thermo-elastic peak frequency & 18\,kHz ($\phi=10^{-6}$)\\

Pendulum Q factor & $2 \times 10^6$ \\

Suspension control & Eddy current damping, coil-magnet actuators and electro 
static drives\\

\hline

  Interferometer configuration & Zero area 
Sagnac interferometer with high-finesse triangular arm cavities\\

Arm cavity length (round trip) & $1.3$\,m ($\approx 2.83$\,m)\\

Arm cavity input mirrors & Fused silica, $T=500$\,ppm\\

Dielectric mirror coatings & Silica/Tantala double stacks 
($\phi_{\tiny{\textnormal{SiO}_2}}=4\times 10^{-5}$, 
$\phi_{\tiny{\textnormal{Ta}_2\textnormal{O}_5}}=2.3 \times 10^{-4}$)\\

Arm cavity finesse & $\sim$ 10000\\

Optical power & 0.3\,W at beam splitter, $\sim$ 1\,kW in the arms\\

Beam radius on cavity mirrors & 1.12\,mm (input mirror); 1.01\,mm (end 
mirrors)\\

Arm cavity round trip loss & $\leq$ 25\,ppm\\

\hline

Interferometric sensing of auxiliary degrees of freedom & Heterodyne signal 
readout\\

Main interferometer readout & Balanced homodyne detection with suspended 
optical local 
oscillator path\\

Readout quantum efficiency & $\geq 95$\,\%\\
 \hline
\end{tabular}
\end{table}
The plots in figure \ref{fig:SSM_MI_noise_budgets_combined_25ppm_loss} show 
the noise model of our SSM experiment in comparison to an equivalent Michelson 
interferometer, including quantum noise \cite{DanKhaLRR2012, 
Danilishin2014}, thermal noise of the mirror suspensions \cite{Pitkin11, 
Penn06, Alshourbagy06, Gossler04, Brif99}, 
Brownian noise \cite{Harry07} and thermo-optic noise \cite{Evans08} of the 
dielectric mirror coatings, Brownian noise of the mirror substrates 
\cite{Bondu98, Liu00} and optical pathlength noise \cite{ZuckerWhitcomb96} as 
well as Brownian force noise due to residual gas in the vacuum envelope 
\cite{Cavalleri10, Dolesi11}. The code architecture of our noise modelling 
software is based on the interferometer noise modelling tool 
GWINC \cite{LIGO-GWINC}. 


Loss is the enemy of any optical experiment probing at effects which are 
governed by quantum mechanics. Theory predicts that the net reduction of QRPN 
in 
a Sagnac speed meter critically 
depends on the incurred optical losses \cite{Danilishin2004}, in particular on
round trip loss inside the arm cavities. Minimal speed meter-type 
QRPN can only be 
achieved in an interferometer with 
perfect, lossless optics. For our experiment we are aiming for a round trip 
optical loss of equal or less than 25\,ppm in each arm cavity and a detection 
efficiency of the main interferometer signal of better than 95\,\%. Our 
estimate of the round trip loss in the arm cavities is based on 
recent investigations of the feasibility of low loss filter cavities for the 
generation of frequency dependent squeezed states of light \cite{Evans13, 
Isogai13} for the broadband quantum enhancement of GW detectors. For the chosen 
value of 25\,ppm per round trip it was taken into 
account that the arm cavities in our experiment will be triangular ones, formed 
by three mirrors each, {\em cf.~}section \ref{sec:realisation}.

\begin{figure}[Ht]
\centering
\includegraphics[width=0.97\textwidth]
{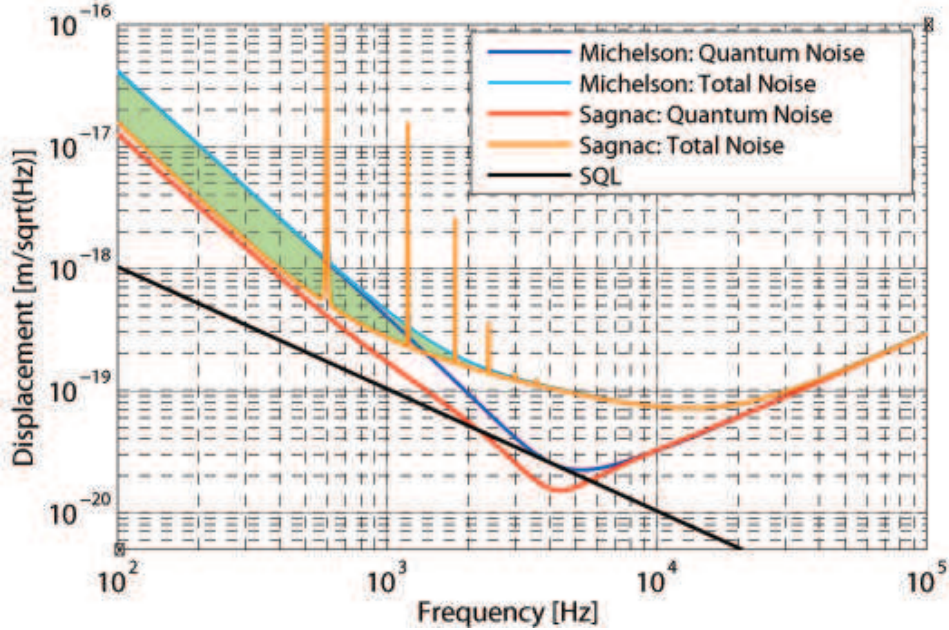}
\caption{An illustration of the higher sensitivity provided by the 
Sagnac speed meter in comparison to a Michelson position meter.  Our 
goal is to demonstrate the improvement represented by the green shaded 
region on the plot.}
\label{fig:QN_and_total_noise_SSM_MI_25ppm}
\end{figure}

The proposed SSM design achieves the goal of a better sensitivity than an 
equivalent Michelson-based 
interferometer at 
frequencies below approximately 3\,kHz. As 
can be seen in figure 
\ref{fig:QN_and_total_noise_SSM_MI_25ppm}, at frequencies of hundreds of Hz the 
Sagnac interferometer outperforms the reference Michelson interferometer by up 
to a factor of $\approx 2.8$. In our experiment we are aiming to reach a 
sensitivity within the green shaded region in figure 
\ref{fig:QN_and_total_noise_SSM_MI_25ppm}, to demonstrate the reduction of QRPN 
in a Sagnac speed meter. If we assume a slightly lower round trip loss of 
20\,ppm per cavity the gap between the total noise in the Sagnac and Michelson 
interferometer increases to a factor of $\approx 3$ and for a round trip loss 
of 15\,ppm per cavity the gap increases to a factor of $\approx 3.2$. For a 
slightly higher round trip loss of 30\,ppm per cavity the gap is reduced to a 
factor of $\approx 2.6$.
Note that further increased margin could be provided by reducing 
coating Brownian noise, which dominates in the measurement band at 
frequencies 
above approximately 800\,Hz, as can be seen in figure 
\ref{fig:SSM_MI_noise_budgets_combined_25ppm_loss}. Crystalline coatings were 
reported to be capable 
of reducing this noise by a factor of $\sim 3$ \cite{Cole2013}. Although a 
further reduction of this noise would increase the signal-to-noise ratio in our 
experiment, it is not crucial to reach our main objective of demonstrating 
reduced quantum noise at low frequencies.

Once the reduction of quantum back-action noise has been 
demonstrated it is planned to carry out a detailed experimental study of the 
influence of optical losses and asymmetries between the arm cavities on the 
performance of the interferometer. The experimental validation of the 
predictions made in theoretical models will contribute substantially to the 
understanding of Sagnac-based interferometer configuration and to laying a 
foundation for establishing Sagnac speed meters as baseline configurations 
for future generation GW detectors.

\section{Technical realisation of the Sagnac 
Speed Meter experiment}\label{sec:realisation}
In this section we present the design of the SSM experiment testbed 
as well as the optical layout of the core interferometer.

The SSM experiment is currently being set up in the Glasgow interferometer 
prototype laboratory. The facility was designed to meet class 1000 clean room 
standards. However, measurements show that a better performance is achieved in 
practice. To suppress acoustic coupling, residual gas noise and refractive 
index fluctuations, the whole interferometer will be operated in vacuum. The 
vacuum system is enclosed in additional clean tents to provide a clean 
environment for the SSM experiment of class 10 or better, to mitigate the risk 
of increased optical loss due to contamination of the optical elements. The 
experiment will be carried out at room temperature and fused silica will be used 
as the optics' substrate material throughout.

For the SSM experiment, critical classical noise sources are (i) seismically 
driven motion of the test mass mirrors, i.e.~seismic noise, (ii) Brownian 
thermal noise of the mirror suspension 
systems, especially of the suspension fibres, i.e.~suspension thermal noise and 
(iii) Brownian noise of the highly reflective dielectric mirror coatings of the 
core interferometer optics, i.e.~coating Brownian thermal noise.

\begin{figure}[t]
\centering
\includegraphics[width=0.8\textwidth]{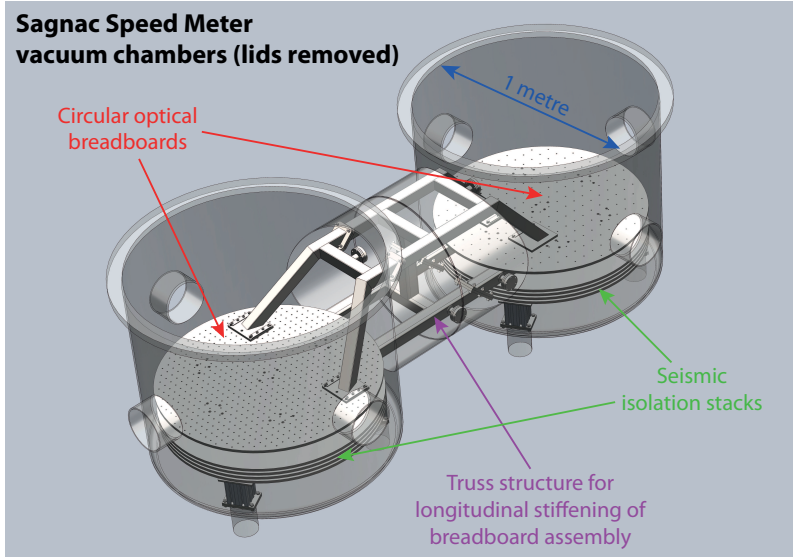}
\caption{Rendered image of the SSM experiment vacuum enclosure, 
containing 4-stage seismic stacks for pre-isolation and the optical 
table structure. For stiffening, to provide rigid motion of the two 
platforms at low frequencies, the circular stainless steel optical tables in 
the two tanks are rigidly connected with a box section truss structure. All 
in-vacuum assemblies are custom made for the SSM experiment.}
\label{fig:cad_drawing_tank_stacks_tables}
\end{figure}

Shielding of the interferometer from seismic motion of the ground is achieved 
with a stack-type pre-isolation system in combination with multi-stage cascaded 
pendulum mirror suspensions. For the seismic isolation platforms, a passive 
multi layer stack consisting of 
alternating layers of fluoroelastomer springs and circular stainless 
steel mass elements was adopted. With our 4-layer stack system we 
aim for vertical resonance frequencies around 18\,Hz and a suppression of 
seismic motion of more than 60\,dB at frequencies above 200\,Hz. 
Additionally, the two circular optical platforms supporting the mirror 
suspensions will 
be connected and stiffened with a truss structure, in order to ensure rigid 
motion of both platforms at low frequencies. A rendered image of the vacuum 
enclosure, seismic pre-isolation system and optical tables is shown in figure 
\ref{fig:cad_drawing_tank_stacks_tables}.

\begin{figure}[th]
\centering
\includegraphics[width=1\textwidth, trim=0cm 1cm 0cm 0cm, 
clip=true,]{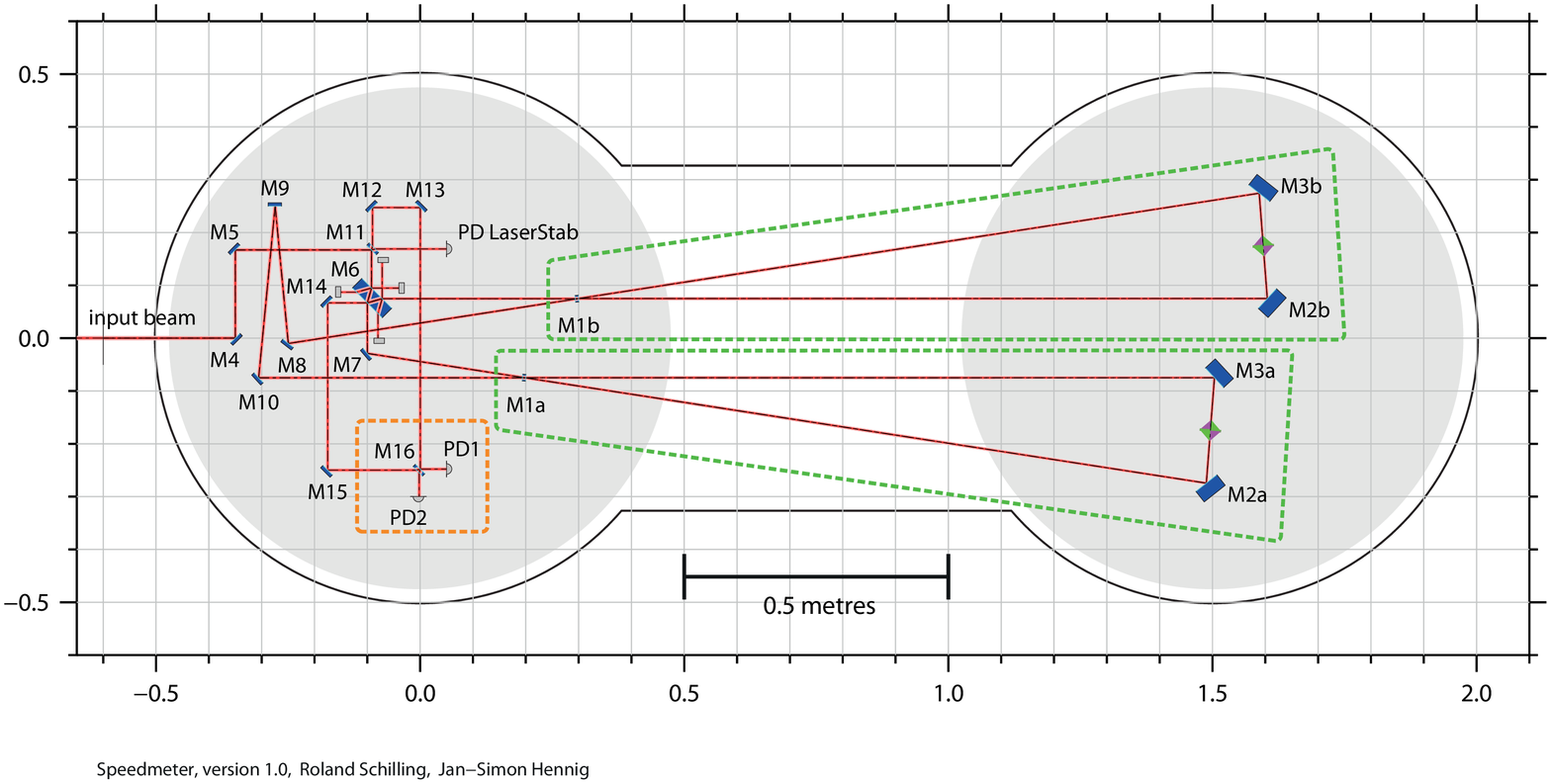}
\caption{Optical layout of 
the Glasgow SSM experiment. To isolate the experiment from seismic as well as 
environmental noise, the core interferometer and signal detection optics will 
be 
housed inside a vacuum enclosure. All optical elements shown in the drawing 
will 
be suspended  with cascaded pendulum suspension systems. These will be mounted 
on two rigidly connected circular optical tables which form the uppermost stage 
of a multi-layers passive isolation stack. The laser beam enters the left 
chamber from outside vacuum by means of free beam coupling. Whereas the left 
vacuum chamber will contain the majority of the optical elements, including 
steering optics, balanced homodyne detection (orange dashed frame), the central 
Sagnac beam splitter (labelled as M6) as well as the arm cavity input couplers 
(M1a, M1b), the chamber on the right hand side of the sketch will only contain 
the highly reflective beam directors (M2a/b, M3a/b) of the two triangular arm 
cavities (green dashed frames). The image was generated with \textsc{OptoCad} 
\cite{OptoCad}. The diamond-shaped symbols between the mirrors M2a/b and 
M3a/b indicate the position of the beam waist of the arm cavity eigenmodes.}
\label{fig:optocad_layout}
\end{figure}

All mirrors inside the vacuum system, including the main beam splitter of the 
Sagnac interferometer, the arm cavity mirrors as well as numerous steering 
mirrors, will be suspended with cascaded pendulum suspensions. An overview of 
the in-vacuum optical layout is shown in figure \ref{fig:optocad_layout}.
The designs of the suspensions used in our SSM experiment are conceptually 
similar to the designs used in large scale GW interferometers \cite{Plissi2000} 
and range from simple metal wire suspensions of the steering mirrors to triple 
pendulum suspensions with an all-monolithic final stage for input and end 
mirrors of the arm cavities. However, low mass mirrors 
substantially complicate the task of designing low-dissipation suspensions, 
e.g.~due to a higher significance of surface loss as a consequence of a high 
surface-to-volume ratio on the one hand and cross 
couplings due to the smaller scale and the effects of fabrication tolerances on 
the other hand.

With these suspensions we aim for lower thermal noise than was observed in 
foregoing experiments with similar mirror masses. A key element of the core 
optics suspensions will be $10\,\mu$m -- $20\,\mu$m diameter fused silica 
fibres, which have recently become feasible to be manufactured due to advances 
in precision controlled fibre production at Glasgow. It is planned to utilise 
the fibre pulling and welding techniques and infrastructure developed for the 
Advanced LIGO monolithic suspensions \cite{Heptonstall2011, Bell2014}, albeit 
on a much smaller scale. For our noise model a value of 
$Q_{\textnormal{\small{pend}}}=2\times 10^6$ was assumed for the pendulum 
quality factor which results in sufficiently low suspension thermal noise in 
the SSM experiment, \emph{cf.~}section \ref{sec:idea}. The stress of the fibres 
for the one gram mirror 
suspension was calculated to be approximately a factor of 500 below the 
breaking strength of fused silica.

The optical layout of the SSM experiment needs 
to fulfill two crucial 
requirements to ensure compatibility with the target sensitivity. First, 
the influence of coating Brownian thermal noise needs to be reduced below 
quantum noise level, which will be achieved with state-of-the-art Titania-doped 
SiO$_2$/Ta$_2$O$_5$ multilayer coatings \cite{Harry07}, in combination with 
large laser beam spots on the highly reflective test mass surfaces 
\cite{Lovelace07}. Second, the optical setup needs to be designed to ensure 
sufficient stability of the optical cavities with gram-scale mirrors at 
circulating powers in the kW range. Whereas for initial experiments it is aimed 
for optical powers in the arm cavities of approximately one kW, this value may 
be increased towards several kW in subsequent experiments.

For the SSM experiment we have chosen to employ triangular arm cavities as a 
straightforward approach to separating the incoming from the outgoing laser beam 
at the cavity input mirrors, {\em cf.~}figure \ref{fig:optocad_layout}. 
Alternatively, beam separation can be achieved e.g.~by making use of different 
polarisation states of the fields incident and returning from the arm cavities 
\cite{Beyersdorf1999}, which is traditionally referred to as a ``polarisation 
Sagnac'' configuration and has been proposed as a candidate QND configuration 
for Advanced LIGO \cite{Danilishin2004} and for the Einstein Telescope GW 
observatory \cite{Wang2013}. However, despite potentially increased cavity round 
trip loss in the configuration with triangular arm cavities, due to the 
increased number of mirrors per cavity by one, the polarisation Sagnac 
configuration was not chosen for our experiment to avoid the risk of performance 
limitations associated with limited extinction of polarising optics.

\begin{figure}[t]
\centering
\includegraphics[width=1\textwidth, trim=0cm 0cm 0cm 0cm, 
clip=true,]{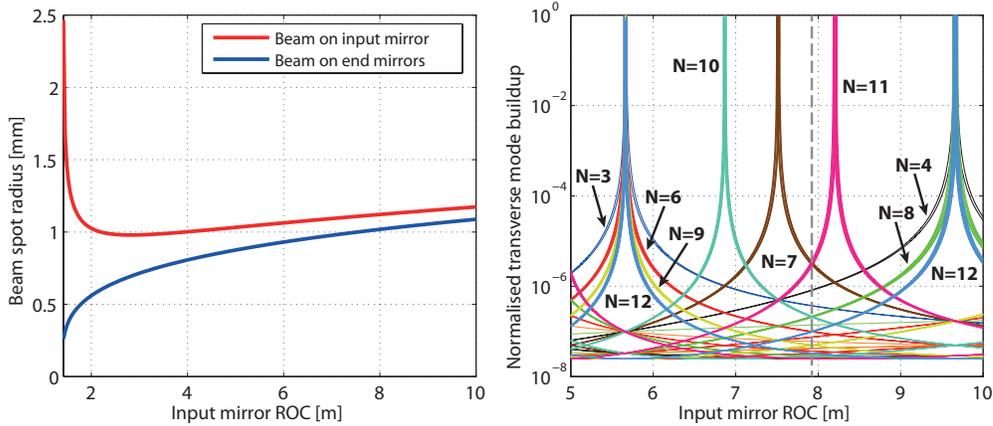}
\caption{\textbf{Left plot:} Beam spot radii on the SSM arm cavity mirrors as a 
function of the input mirror ROC. A round trip length of 2.83\,m 
was assumed. Curvature radii larger than 7.55\,m result in beam 
spots larger than 1\,mm on all three cavity mirrors. \textbf{Right plot:} 
Normalised power buildup, relative to the fundamental mode, of higher order 
transverse modes as a function of the arm cavity input mirror ROC. Higher order 
modes up to 12th order were taken into account. Each set of HOMs of the same 
mode order is represented by a particular colour in the plot. The resonance 
peaks appear to be slightly smeared out due to the non-degeneracy 
of HOMs of the same order in the triangular cavities, resulting in a frequency 
splitting effect. The vertical 
dashed line indicates our chosen arm cavity input mirror ROC of 7.91\,m.
The arm cavities were 
modelled with the interferometer simulation \textsc{Finesse} \cite{Freise04}, 
assuming infinite mirror apertures. 
}
\label{fig:cavity_design}
\end{figure}

Each of the two arm cavities will be formed by a curved one gram input mirror 
(M1a/b in figure \ref{fig:optocad_layout}) and two planar 100\,g end mirrors 
(M2a/b, M3a/b) with diameters of $\approx 10$\,mm and $\approx 50$\,mm, 
respectively. Thus, for a given round 
trip length of the cavities the eigenmode geometry is fully determined by the 
radius of curvature (ROC) of the input mirror. The dependence of beam 
spot radii on the cavity mirrors and the buildup of higher order transverse 
modes (HOMs) on the input mirror ROC is illustrated in figure 
\ref{fig:cavity_design}. For the input mirror ROC a value of 7.91\,m 
was chosen which leads to beam spot radii $\geq 1$\,mm on 
all cavity mirrors. Assuming a clear aperture radius of 
the arm cavity input mirrors of $\approx 4$\,mm and beam spot radii of $\approx 
1.12$\,mm on the input mirrors, \emph{cf.~}left plot in figure \ref{fig:cavity_design}, HOMs up 
to approximately 12th order will 
resonate inside the cavities with relatively low losses \cite{Barriga2007}. 
Our chosen value for the input mirror ROC results in a comparatively low 
resonant buildup of low order HOMs in the arm cavities, {\em cf.~}right plot in 
figure \ref{fig:cavity_design}, which could otherwise 
limit the performance of the experiment. Based on the generalised approach 
detailed in \cite{SiegmanLasers}, our triangular arm cavity design 
exhibits geometric stabilities of $m^2_t \approx 0.411$ in the tangential 
plane and $m^2_s \approx 0.414$ in the sagittal plane, which were obtained 
from the propagation matrices describing one full cavity round 
trip.

Arranging for beam spots larger by a factor of $\approx 2$ would require an 
input mirror ROC of approximately 100\,m which came at the expense of a 
strongly reduced geometric stability and potentially lower robustness to 
misalignment of the cavities. Furthermore, taking the free aperture radius of 
the arm cavity input mirrors of $\approx 4$\,mm into account, beam spots with a 
radius of 2\,mm or larger are prone to substantially increased scattering and 
diffraction loss. To mitigate these losses the diameter of the 
arm cavity input mirrors could be increased. However, to maintain the same 
ratio of beam spot radius to clear aperture radius, while also maintaining the 
same substrate aspect ratio optimised for low thermal noise \cite{Somiya2009}, 
an increase of the beam spot radius by a factor of two would result in an 
8-fold increase of the mirror mass. As a consequence this would require a 
substantial increase of the circulating laser power to reach the same level of 
quantum back-action noise as in the configuration with smaller beam spots.

Further important drivers for the design of the optical layout of our 
experiment are the requirement to ensure optimal mode matching of the two arm 
cavities, sufficient means of optimising the angular alignment of the 
interferometer as well as optimal beam overlap at the balanced homodyne 
detector 
({\em cf.~}section \ref{sec:challenges}), to keep optical detection losses at a 
minimum. For the planned geometry of the triangular SSM arm cavities, 
calculations based on the model presented in \cite{T920004} have shown that the 
intrinsic astigmatism results in a degradation of the theoretically achievable 
mode matching efficiency in the sub-ppm range. This result was confirmed in 
numerical simulations of the arm cavities. Hence, no significant impact on the 
balanced homodyne detector contrast is expected. However, if the signal 
field leaving the interferometer turned out to be more astigmatic than 
expected, an astigmatic LO beam could be employed to recover the homodyne 
contrast.

The input laser beam, coming from the left in the drawing in figure 
\ref{fig:optocad_layout}, is directed into the 
vacuum tank and propagates towards the main 50/50 beam splitter (denoted M6) 
where the beam is split equally and guided into the two arm cavities. Mode 
matching to the arm cavities with a beam waist located between the far mirrors 
M2a, M3a and M2b, M3b, respectively, is achieved with a mode matching telescope 
outside vacuum. The cavities will be mode matched among each other with the aid 
of a curved steering mirror labelled M9 in the drawing. For this mirror, a 
small 
angle of incidence was chosen to avoid distortion of the reflected beam. For 
beam steering and auxiliary optics, mirrors with a diameter of 30\,mm 
will 
be used. For the main beam splitter, M6, a diameter of 80\,mm was chosen, to 
allow 
better control of AR and secondary reflections.

The bright port of the Sagnac interferometer was chosen as the source of the 
local oscillator (LO) field for balanced homodyne detection. In contrast to 
picking off a fraction of the input beam or, alternatively, light reflected 
from the AR coating of the main beam splitter \cite{Fritschel2014}, the beam 
exiting the bright port of the interferometer is easily accessible and exhibits 
the same wave front curvature as the signal beam, i.e.~the two fields can be 
superimposed without the need for additional focussing elements in one of the 
beam paths. Longitudinal and angular degrees of freedom of the LO field and the 
signal field can be actuated upon with the mirrors M12, M13 and M14, M15, 
respectively. Approximately 10\,mW of laser light are planned to be picked off 
for the LO beam. However, if requirements change in the course of our 
experiment, the LO power can be adjusted 
by swapping mirrors in the optical layout with ones with different 
reflectivity. Besides the mirror M11, whose transmittance also determines the 
laser power reaching the in-vacuum photo diode for laser amplitude 
stabilisation, the transmittances of the mirrors M12 and M13 in the LO 
path can be adjusted to provide independence of the LO power from the 
power of the light picked off for laser stabilisation.

Similar to the Michelson-based interferometers used in GW detection our SSM 
interferometer will require feedback control of various longitudinal and angular 
degrees of freedom in the optical setup, to ensure maximisation of 
resonant light buildup and minimisation of noise couplings to the signal 
port of the interferometer. We are planning to take the same technical approach 
currently employed in  Michelson-based large and small scale interferometers, 
i.e. heterodyne signal 
extraction based on frontal phase modulation in conjunction with the 
well-known Pound-Drever-Hall technique \cite{PDH}. To allow for flexibility in 
the 
control scheme of the interferometer it is planned to include actuators in all 
core optics suspensions. Coil-magnet type actuators in combination with 
electrostatic drives will serve to actuate on the relevant longitudinal and 
angular 
degrees of freedom of the core optics. A new type of 
electro-static actuator, similar to the design used in the AEI 10\,m Prototype 
\cite{Wittel2014}, is planned to be used for direct actuation on the low mass 
arm cavity input mirrors. Longitudinal and angular control is planned for the 
two arm cavities, the balanced homodyne detector and, if needed, also for the 
optical path lengths in the central Sagnac interferometer and the path of the 
beams travelling from the first to the second arm cavity. The main calibrated 
output signal in the SSM experiment will be obtained from the balanced homodyne 
detector. Detailed numerical investigations to inform the final interferometric 
sensing and controls system design are currently underway.

\section{Experimental challenges}\label{sec:challenges}
In many respects, the SSM experiment will require 
existing and well-established interferometry concepts to be pushed to their 
limits and, moreover, it will be necessary to adopt techniques which have not 
yet been practically implemented and tested under comparable circumstances. In 
this section we discuss some experimental challenges which we expect to face in 
the course of the construction and operation of the SSM experiment.

\subsection{Radiation pressure-dominated test mass dynamics}
The SSM will be operated in a regime in which the test mass dynamics are
dominated by radiation pressure effects. The combination of light 
mirrors and moderately high circulating optical power is expected to 
drive longitudinal and angular optical springs, formed due to 
radiation pressure during lock acquisition and in the locked interferometer, to 
higher frequencies than e.g.~in full-scale GW detectors.  This will require the 
development of new lock acquisition strategies and improved control schemes to 
keep the interferometer optics at their designated operating points.

Longitudinal optical springs occur in detuned cavities and change the dynamics 
of the experiment by opto-mechanical coupling. To maintain control over the 
affected resonators, actuators with sufficiently large actuation range are 
required. With multi-stage suspensions it would be preferable to 
actuate on one of the upper masses, thus benefiting from the filtering of 
actuator noise by the pendulum stages. This, on the other hand implies that 
actuation only at low frequencies will be possible, due to the low pass filter 
characteristic of the pendulum stages. Optical springs in the kHz range may 
require direct actuation on the test mass for compensation which would require 
the use of very low noise actuators. However, the occurrence of optical springs 
during lock acquisition may be mitigated with the aid of a special tailored 
lock acquisition scheme, e.g.~the transition of the experiment to the targeted 
high-power state of operation after lock acquisition at lower optical power. 

For the effect of radiation pressure-induced angular optical springs it was 
shown that these can be mitigated by adopting a suitable arm cavity design 
\cite{Sidles06} and by means of appropriate angular feedback 
control \cite{Dooley13}. The latter approach, however, is known to be 
challenging to realise for angular springs ocurring at high frequencies due to 
large required control bandwidths. Theoretical investigations of angular 
instabilities in the triangular cavities in the LIGO interferometers showed 
that for a geometrically stable cavity the angular modes in the 
horizontal plane, i.e.~in yaw, are always intrinsically stable \cite{T030275}. 
This property was confirmed experimentally in \cite{Matsumoto2014}. In the same 
article the authors propose a new approach to reduce suspension thermal noise 
in interferometric weak force measurements which is based on using triangular 
cavities, which do not require angular control in yaw due to the aforementioned 
effect, and single-wire mirror suspensions. Although this approach may offer 
reduced low frequency thermal noise, for the SSM experiment the standard 
approach offering the possibility of full angular control of the mirrors will 
be taken, to avoid relying on new, less mature techniques where 
possible.

\subsection{Main interferometer signal readout}

The fact that in 
a Sagnac interferometer the light returning from the arms always 
interferes destructively at the signal port, regardless of any microscopic 
offsets, has implications on the effectiveness of sensing and control 
schemes. On the one hand, this property reduces, by one, the number of length 
degrees of 
freedom in the interferometer that require feedback control to keep the 
instrument on its nominal operating point \cite{Sun96, Mizuno97}. On the other 
hand, this effect 
renders the DC readout scheme, which is the standard method of reading 
out the main interferometer signal in second generation GW detectors 
\cite{Ward08, Hild2009, Fricke2012}, unsuitable for our experiment. For this 
reason it is planned to utilise balanced homodyne detection in the 
SSM experiment, {\em cf.~}section \ref{sec:idea}. 

In balanced homodyne detection, other than in the DC 
readout scheme, an external optical LO field is superimposed on 
a beam splitter with the field exiting the signal port of the interferometer, 
which allows the selection of an arbitrary readout quadrature by tuning the 
relative phase of the LO and signal field accordingly \cite{Bachor04}. It is an 
intrinsic property of balanced homodyne detectors that common-mode noise is 
suppressed. By careful fine tuning of the optical and electronic balancing, 
common mode rejection ratios of up to approximately 80\,dB have been 
consistently achieved \cite{Stefszky12}. Even though balanced 
homodyne detection is an established and well-understood tool in bench-top 
quantum optics experiments, it has not yet been employed for reading out 
suspended interferometers at audio frequencies. It is expected that the 
frequency stability of the LO field will be crucial to read out displacement 
noise down to the $10^{-19}$\,m/$\sqrt{\textnormal{Hz}}$ level. 

Also, carrier light leaking from within the interferometer into the 
detection port may give rise to additional readout noise due to a 
\emph{reverse homodyning} effect, by acting as an unwanted second local 
oscillator at the 
balanced homodyne detector. In the SSM experiment the level of carrier leakage 
will strongly depend e.g.~on the beam splitting ratio of the main Sagnac beam 
splitter. Since any attempt of attenuating the beam exiting the interferometer 
towards the balanced homodyne detector will introduce additional optical losses, 
it will be necessary to control carrier leakage by formulating sufficiently 
stringent tolerances for the relevant core optics parameters.

Detailed investigations of aspects such as the impact of LO noise and carrier 
light leakage on the performance of the planned balanced homodyne detection 
scheme are currently ongoing, experimentally as well as theoretically. 
Optionally, heterodyne readout schemes which are less susceptible 
to low-frequency noise may be considered as alternatives.

\section{Summary and Outlook}\label{sec:summary}

In this article we have presented a conceptual design for a speed meter 
proof-of-principle experiment. The construction of the experiment is well under 
way, with the vacuum system and the seismic pre-isolation completed. Next steps 
will include, amongst others, a detailed analysis of the relation of  
asymmetric optical losses in the various sections of the interferometer and 
their influence on quantum noise \cite{Danilishin2014}, the technical 
design of the multi-stage suspension system for the one gram mirrors as well as 
the design of a longitudinal and angular sensing and control scheme for the 
interferometer, including detailed control noise studies.
 
The experimental demonstration of the reduction of
quantum back-action noise in a Sagnac speed meter interferometer is expected 
to have significant impact on the field of ultra-high precision interferometery 
and will be the first step towards qualifying Sagnac topologies as baseline 
interferometer designs for future generation GW observatories such as the 
Einstein Telescope.


\ack{The work described in this article is funded by the
European Research Council (ERC-2012-StG: 307245).
We are grateful for support from Science and
Technology Facilities Council (STFC), the Humboldt
Foundation, the International Max Planck Partnership
(IMPP) and the ASPERA ET-R\&D project.}

\end{document}